\documentclass{PoS}
\usepackage{enumerate}
\usepackage{amssymb}
\usepackage{amsmath}

\newcommand{\beqn}{\begin{eqnarray}}
\newcommand{\eeqn}{\end{eqnarray}}
\newcommand{\be}{\begin{equation}}
\newcommand{\ee}{\end{equation}}

\newcommand{\ba}{\begin{array}{c}}
\newcommand{\bat}{\begin{array}{cc}}
\newcommand{\ea}{\end{array}}

\newcommand{\bi}{\begin{itemize}}
\newcommand{\ei}{\end{itemize}}

\newcommand{\ket}{\,\rangle}
\newcommand{\bra}{\langle \,}

\newcommand{\Frac}[2]{\frac{\displaystyle #1}{\displaystyle #2}}

\newcommand{\cO}{{\cal O}}

\newcommand{\mF}{\mathcal{F}}
\newcommand{\mG}{\mathcal{G}}
\newcommand{\mH}{\mathcal{H}}

\newcommand{\mK}{\mathcal{K}}
\newcommand{\mL}{\mathcal{L}}
\newcommand{\mM}{\mathcal{M}}
\newcommand{\mO}{\mathcal{O}}

\newcommand{\Int}{\displaystyle{\int}}

\newcommand{\bear}{\begin{eqnarray}}
\newcommand{\eear}{\end{eqnarray}}

\newcommand{\divergence}{\Frac{\mu^{D-4}}{16\pi^2(D-4)}}
\newcommand{\nn}{\nonumber}

\newcommand{\sla}{\hspace*{-0.2cm}\slash  }

\title{
\vspace*{-5cm}\hspace*{0cm}
{\small
 TUM-HEP-1023/15,
 FTUAM-15-38,
 IFT-UAM/CSIC-15-117
}
\vspace*{3cm}\hspace*{-15cm}
\\
One-loop corrections to the Higgs electroweak chiral Lagrangian}

\ShortTitle{One-loop corrections to the ECLh}

\author{Feng-Kun Guo \\
        State Key Laboratory of Theoretical Physics,
Institute of Theoretical Physics, CAS, Beijing 100190, China  \\
        E-mail: \email{fkguo@itp.ac.cn
}}

\author{Pedro Ruiz-Femen\'\i a \\
        Physik Department T31, James-Franck-Strasse, Technische Universit\"at M\"unchen, D-85748 Garching, Germany \\
        E-mail: \email{pedro.ruiz-femenia@tum.de}}

\author{\speaker{Juan Jos\'e Sanz Cillero}
\thanks{
JJSC would like to thank the organizers for the nice conference
and their help whenever it was needed.
The work of FKG is supported in part by the NSFC and DFG through funds
provided to the Sino-German
CRC 110 ``Symmetries and the Emergence of Structure in QCD"(NSFC Grant No.
11261130311) and
NSFC (Grant No. 11165005)        
and the work of JJSC is supported by ERDF funds from the European Commission
[FPA2010-17747, FPA2013-44773-P, SEV-2012-0249, CSD2007-00042].   
}\\
        Departamento de F\'\i sica Te\'orica and Instituto de F\'\i sica Te\'orica, IFT-UAM/CSIC,
Universidad Aut\'onoma de Madrid, Cantoblanco, 28049 Madrid, Spain\\
        E-mail: \email{juanj.sanz@uam.es}}

\abstract{
In this talk we study beyond Standard Model scenarios where the Higgs is non-linearly realized.
The one-loop ultraviolet divergences of the low-energy effective theory at next-to-leading order, $\cO(p^4)$,
are computed by means of the background-field method and the heat-kernel expansion.
The power counting in non-linear theories shows that these divergences are determined by the leading-order effective Lagrangian $\mL_2$.
We focus our attention on the most important $\cO(p^4)$ divergences,  which
are provided by the loops of Higgs and electroweak Goldstones,
as these particle are the only ones that couple through derivatives in $\mL_2$.
The one-loop divergences are renormalized by $\cO(p^4)$ effective
operators, and set their running. This implies the presence of chiral logarithms in the amplitudes along with the
$\cO(p^4)$ low-energy couplings, which
are of a similar importance and should not be neglected in next-to-leading order effective theory calculations,
e.g. in composite scenarios.
}

\FullConference{18th International Conference From the Planck Scale
  to the Electroweak Scale \\
  25-29 May 2015\\
  Ioannina, Greece }

\begin{document}

\section{Introduction}

In these proceedings we study non-linear electroweak (EW) effective theories including a light Higgs, which we will denote as the electroweak chiral
Lagrangian with a light Higgs (ECLh).
In Ref.~\cite{Guo:2015} we have computed the next-to-leading order (NLO) corrections induced by scalar boson
(Higgs $h$ and EW Goldstones $\omega^a$) one-loop diagrams. These contributions provide the one-loop
corrections to the amplitude that grow with the energy as $p^4$, as these particles
are the only ones that couple derivatively
in the lowest-order (LO) effective Lagrangian~\cite{Longhitano:1980iz,EW-chiral-counting,ECLh-Gavela}.
We used the background field method and heat-kernel expansion
to extract the ultraviolet divergences of the theory at NLO, i.e.,
$\cO(p^4)$, where $p$ is the effective field theory (EFT) expansion parameter and refers to any infrared (IR) scale of the EFT
--external momenta or masses of the particles in the EFT--.
Many beyond Standard Model (BSM) scenarios show this non-linear realization,
which are typically
strongly coupled theories with composite states~\cite{composite-rev}.
A common feature in non-linear EFT's is that one-loop corrections are formally
of the same order as tree-level contributions from higher dimension
operators~\cite{chpt,Weinberg:1979}.
Phenomenologically, these two types of corrections are of a similar size,
provided the scale of non-linearity
$\Lambda_{\rm non-lin}$ that suppresses the $h$ and $\omega^a$ loops and the composite resonance masses $M_R$
are similar~\cite{Cillero-Blois}:
$$
\mbox{NLO tree diagrams} \quad \sim \quad \mbox{NLO loop diagrams .}
$$

\section{Low-energy Lagrangian and chiral counting}

The low-energy theory is given by the usual ingredients~\cite{EFT}:
\begin{itemize}

\item
{\bf EFT particle content:} the singlet Higgs field $h$,
the non-linearly realized triplet of EW Goldstones $\omega^a$  and
the SM gauge bosons and fermions.

\item
{\bf EFT symmetries:} we based our analysis on the symmetry pattern
of the SM scalar sector $\mG=SU(2)_L\times SU(2)_R$,
which breaks down spontaneously into the custodial subgroup $\mH=SU(2)_{L+R}$.
The subgroup $SU(2)_L\times U(1)_Y\in \mG$ is gauged.
~\footnote{
When fermions are included in the theory  $\mG$ must be enlarged to
$\mG= SU(2)_L\times SU(2)_R\times U(1)_{\rm B-L} \supset SU(2)_L\times U(1)_Y$ and
$\mH= SU(2)_{L+R}\times U(1)_{\rm B-L}\supset U(1)_{\rm EM}$~\cite{Hirn:2004},
with ${\rm B}$ and ${\rm L}$ the baryon and lepton numbers, respectively.
}

\item
{\bf Locality:} The underlying theory may contain non-local exchanges of heavy states.
Nevertheless, in the low-energy limit
the effective action is provided by an expansion of local operators.

\end{itemize}

In the case of non-linear Lagrangians the classification of the EFT operators
in terms of their canonical dimension is not appropriate and what really ponders
the importance of an operator
in an observable is their ``chiral'' scaling $p^{\hat{d}}$ in terms of the infrared scales
$p$~\cite{EW-chiral-counting,Weinberg:1979,Hirn:2004,Georgi-Manohar,1loop-AA-scat}.
The ECLh is organized in the form
\bear
\mL_{\rm ECLh} &=& \mL_{2} \, +\, \mL_{4} \, +\, ...
\eear
where the terms of order $p^{\hat{d}}$ have the generic form~\cite{1loop-AA-scat,Santos:2015}
\bear
\mL_{\hat{d}} &\sim  & \sum_{k,n_F}\,
c_{(\hat{d})}\, p^{d} \, \left(\Frac{\chi}{v}\right)^k
\, \left(\Frac{\overline{\psi} \psi }{v^2}\right)^{n_F/2}\, ,
\label{eq.Ld}
\eear
with $\hat{d}= d + n_F/2$, $v=(\sqrt{2} G_F)^{-1/2}=246$~GeV,
$\chi$ ($\psi$) representing any bosonic (fermionic) fields in the ECLh, and
being $p$ any infrared scale appropriately acting on the fields (derivatives,
masses of the particles in the EFT, etc.).
In the lowest order case $\hat{d}=2$ one has couplings $c^{(2)}\sim v^2$.
%
%
The counting can be established more precisely
by further classifying what we mean by $p$ in our operators
(explicit derivatives, fermion masses, etc.)~\cite{EW-chiral-counting}.
Beyond naive dimensional analysis  (NDA),
one usually makes further assumptions on the scaling of the coupling of the composite sectors
and the elementary fermions.
Typically one assumes them to be weakly coupled in order to support the phenomenological observation
$m_\psi\ll 4\pi v\approx 3$~TeV and the moderate size of the Yukawa couplings measured
so far at LHC.~\footnote{
Based on pure NDA the Yukawa operators in the ECLh would be $\cO(p^1)$,
spoiling the chiral power expansion. In order to avoid this, one needs to make the phenomenologically supoorted
assumption that the constants
$\lambda_\psi$ that parametrize the couplings between the elementary fermions and the composite scalars
$h$ and $\omega^a$ are further suppressed, scaling at least like $\lambda_\psi  \sim \cO(p)$ or higher in the
chiral counting~\cite{EW-chiral-counting,Santos:2015}.
}

The LO Lagrangian is given by~\cite{EW-chiral-counting,ECLh-Gavela,ECLh-other,SILH},
\bear
\mL_2 &=& \Frac{v^2}{4} \mF_C \bra u_\mu u^\mu \ket
+\Frac{1}{2} (\partial_\mu h)^2 -
v^2 \, V
+ \mL_{YM} +   i \bar{\psi}\,   D \sla \, \psi \, -\, v^2 \bra  J_{S}\ket \, ,
\label{eq.L2}
\eear
where $\bra ...\ket$
stands for the trace of $2\times 2$ EW tensors, $\mL_{YM}$ is the
Yang--Mills Lagrangian for the gauge fields,
$D_\mu$ is the gauge covariant derivative acting on the fermions,
$V[h/v]$ is the Higgs potential and $J_S$
denotes the Yukawa operators that couple the SM fermions to $h$ and $\omega^a$.
The factors of $v$ in the normalization of some terms are  introduced for later convenience.
$\mF_C,\, V$ and $J_S$ are functionals of $x=h/v$, and
have Taylor expansions,
$\mF_C[x]  = 1 + 2 a x + b x^2 +...$,
$J_S[x]=\sum_n J_S^{(n)} x^n/n!$ and
$V[x]= m_h^2 \left( \frac{1}{2}  x^2+\Frac{1}{2} d_3 x^3 + \Frac{1}{8} d_4 x^4 \, +  ... \right)$,
given in terms of the constants
$a,b,m_h$, etc.~\cite{EW-chiral-counting,ECLh-Gavela,ECLh-other,SILH}.
In the non-linear realization of the spontaneous EW symmetry breaking,
the Goldstones are parameterized through the coordinates  $(u_L,u_R)$ of the
$SU(2)_L\times SU(2)_R/SU(2)_{L+R}$ coset space~\cite{CCWZ}, with the $SU(2)$ matrices $u_{L,R}$
being functions of the Goldstone fields $\omega^a$  which enter through the building blocks
\bear
&&u^{}_\mu =
i u_R^\dagger (\partial^{}_\mu-i r^{}_\mu) u^{}_R - iu_L^\dagger (\partial^{}_\mu-i \ell^{}_\mu) u^{}_L\, ,
\quad
\Gamma^{}_\mu = \frac{1}{2} u_R^\dagger (\partial^{}_\mu-i r^{}_\mu) u^{}_R
+\frac{1}{2}u_L^\dagger (\partial^{}_\mu-i \ell^{}_\mu) u^{}_L\, ,
\nn\\
&&\nabla_\mu\, \cdot  =  \partial_\mu \,\cdot \, +\, [\Gamma_\mu , \,\cdot\, ] \, ,
\quad f_\pm^{\mu\nu}
= u_L^\dagger \ell^{\mu\nu} u_L \pm u_R^\dagger r^{\mu\nu} u^{}_R\, ,
\quad
r_{\mu\nu}
= \partial_\mu r_\nu -\partial_\nu r_\mu - i [r_\mu, r_\nu]\, ,
\quad (R\leftrightarrow L)\, ,
\nn\\
&&  J^{}_{S} = J^{}_{YRL} +J_{YRL}^\dag,
  \qquad
  J^{}_{P} = i ( J^{}_{YRL} - J_{YRL}^\dag),
\qquad
  J^{}_{YRL} = - \Frac{1}{\sqrt{2} v} u_R^\dagger  \hat{Y}
\psi^{\alpha}_R \bar{\psi}^{\alpha}_L
u^{}_L   ,
\eear
with $\psi_{R,L}=\frac12(1\pm \gamma^5)\psi$ and the $SU(2)$ doublet $\psi = (t,b)^T$.
The summation over the Dirac index $\alpha$ in
$\psi^{\alpha}_{R\,\, m}  \bar{\psi}^{\alpha}_{L\,\, n}
= -\bar{\psi}^{\alpha}_{L\,\, n}  \psi^{\alpha}_{R\,\, m}  $
is assumed and its tensor structure under $\mG$ and indices $m$ and $n$ are left implicit.
The $2\times 2$ matrix
$\hat{Y}[h/v]$ is a spurion auxiliary field, functional of $h/v$,
which incorporates the fermionic Yukawa coupling~\cite{EW-chiral-counting,Hirn:2004,flavor-ECLh}.
Other SM fermion doublets and the
flavour symmetry breaking between generations can be incorporated
by adding in $J_{YRL}$ an additional family index in the fermion fields,
$\psi^A$, and promoting $\hat{Y}$ to a tensor $\hat{Y}^{AB}$
in the generation space~\cite{flavor-ECLh}.
In our analysis, $\ell_\mu,\, r_\mu , \, \hat{Y}$ are spurion
auxiliary background fields
that keep the invariance of the ECLh action under $\mG$.
When evaluating physical matrix elements, custodial symmetry is then explicitly broken in the same way as in the SM,
keeping only the gauge invariance under the subgroup
$SU(2)_L\times U(1)_Y\subset \mG$~\cite{Longhitano:1980iz,EW-chiral-counting,ECLh-Gavela,ECLh-other},
with the auxiliary fields taking the value
$
\ell_\mu = - \Frac{g}{2} W_\mu^a \sigma^a\, ,
\,\,\,
r_\mu = - \Frac{g'}{2} B_\mu \sigma^3\, ,
\,\,\,
\hat{Y}[h/v]\,  =\,   \hat{y}_t[h/v]\,  P_+\, +\, \hat{y}_b[h/v]\,  P_- $,
with $P_\pm =(1\pm \sigma^3)/2$.

In order to compute the one-loop fluctuations we will consider the coset representatives
$u_L=u_R^\dagger = u$, often expressed in the exponential
parametrization $U=u^2=\exp\{ i \omega^a \sigma^a/v\}$.~\footnote{
Other Goldstone parametrizations are discussed in~\cite{chpt,1loop-AA-scat,1loop-WW-scat-Dobado,1loop-Gavela},
being all of them fully equivalent when describing on-shell matrix elements.
}

The IR scales in the low-energy theory are
\bear
\partial_\mu, \quad  r_\mu,  \quad  \ell_\mu, \quad  m  , \quad g^{(')} v , \quad \hat{Y} v
\quad \sim\quad  \cO(p)\, ,
\eear
with $m=m_{h,W,Z,\psi}$.
Accordingly, covariant derivatives scale as the ordinary ones~\cite{chpt} and the Lagrangian is invariant under $\mG$
at every order in the chiral expansion.
Based on this we are going to sort out the building blocks and
operators according to the assignment~\cite{Weinberg:1979,Georgi-Manohar},
\bear
\Frac{\chi}{v} \quad &\sim&  \quad \cO(p^0)
\qquad\qquad (\mbox{for the boson fields } \chi=h ,\, \omega^a, \, W_\mu^a,\,  B_\mu )\, ,
\nn\\
\Frac{\psi}{v} \quad &\sim& \quad  \cO(p^{1/2})
\qquad\quad (\mbox{for the fermion fields } \psi=t ,\, b, \, \mbox{etc})\, .
\eear
Therefore, the chiral order of the various building blocks reads
\bear
\mF_C  \,\, \sim \,\,  \cO(p^0)\, ,
\qquad u_\mu ,\, \nabla_\mu  \,\, \sim \,\,  \cO(p^1)\, ,
\qquad r_{\mu\nu},\, \ell_{\mu\nu},\, f_\pm^{\mu\nu},\, J_{YRL},\, J_S ,\, J_P,\, V \,\, \sim \,\,  \cO(p^2)\, .
\eear
Hence, the Lagrangian in Eq.~(\ref{eq.L2}) is $\cO(p^2)$ and provides the LO.
%

So far all we did was to sort out the possible terms of the EFT Lagrangian assigning them an order.
The relevance of this classification is that, at low energies,
when one computes the contributions to a given process
the more important ones are given by the Lagrangian operators
with a lower chiral dimension. An arbitrary diagram with vertices from $\mL_{\hat{d}}$
behaves at low energies like~\cite{EW-chiral-counting,Weinberg:1979,Georgi-Manohar,1loop-AA-scat,Santos:2015}
\bear
\mM &\sim& \Frac{p^2}{v^{E -2} } \, \left(\Frac{p^2}{16\pi^2 v^2}\right)^L
\,\prod_{\hat{d} } \left(\Frac{
c_{(\hat{d})}   p^{\hat{d}-2}}{v^2}\right)^{N_{\hat{d}}}\, ,
\label{eq.amp-scaling}
\eear
with the IR scales $p$, $L$ the number of loops,
$N_{\hat{d}}$ the number of subleading vertices from
$\mL_{\hat{d}>2}$ (with coupling $c_{(\hat{d})}$) and $E$ the number
of external legs.
We can have an arbitrary number of $\mL_2$ vertices in the diagram and the amplitude will still have the same scaling
with $p$, provided the number of loops is fixed.
If we add vertices of a higher chiral dimension we will
increase the scaling of the diagram with $p$.
Thus, we have that the one-loop corrections with only $\mL_2$ vertices are $\cO(p^4)$
and their UV divergences are cancelled out by tree-level diagrams
that contain one $\mL_4$ vertex: the renormalization of the effective couplings at $\cO(p^4)$
will render the effective action finite at NLO.

%

\section{Fluctuations around a background field   }

We are going to consider perturbations $\eta$ in the fields around their equations of motion (EoM) solutions.
The Lagrangian in the integrand of the generating functional
has also a corresponding expansion in the perturbation,
where each order in $\eta$ contains relevant information~\cite{heat-kernel}:
\bear
\mL &=& \quad
\underbrace{  \mL^{\cO(\eta^0)} }_{\mbox{Tree-level}} \quad +\quad
\underbrace{  \mL^{\cO(\eta^1)} }_{\mbox{EoM}} \quad +\quad
\underbrace{  \mL^{\cO(\eta^2)} }_{\mbox{1-loop}} \quad +\quad
\underbrace{  \cO(\eta^3)   }_{\mbox{Higher loops}} \, .
\eear
The Lagrangian evaluated at the classical solution provides the tree-level contributions to the effective action,
the requirement that the linear term in the $\eta$ expansion vanishes provides the EoM,
the quadratic fluctuation in $\eta$ provides the 1-loop contributions to the effective action
and higher loops are encoded in the remaining terms of the $\eta$ expansion.

In our analysis~\cite{Guo:2015} we were interested in the one-loop UV divergences
at $\cO(p^4)$,
that is those coming from diagrams with $\mL_2$ vertices. We studied
the structures that grow faster with the energy at one-loop, as $(\rm Energy)^4$.
They were given by
the loops of scalars ($h$ and $\omega^a$) which are the only ones
that couple derivatively in $\mL_2$.
We made the Goldstone representative choice $u_L=u_R^\dagger$~\cite{Pich:2013}
and performed fluctuations of the scalar fields (Higgs and Goldstones)
around the classical background fields  $\bar{h}$ and
$\bar{u}_{L,R}$~\cite{Guo:2015}:
\bear
u_{R,L}&=& \bar{u}_{R,L} \, \exp\left\{ \pm  i \mF_C^{-1/2}\,     \Delta/(2 v)     \right\}\, ,
\qquad\qquad
h = \bar{h}\, +\, \epsilon\, ,
\eear
with $\Delta=\Delta^a \sigma^a$. Without any loss of generality we
introduced the factor $\mF_C^{-1/2}$
in the exponent for later convenience, allowing us to write down the second-order fluctuation of the action
in the canonical form~\cite{heat-kernel}.
To obtain the one-loop effective action within the background field method
we then retained  the quantum fluctuations $\vec{\eta}^T=(\Delta^a, \epsilon)$
up to quadratic order~\cite{heat-kernel}.

Since we are interested in the loops  with only $\mL_2$ vertices, we study the $\eta$ expansion of the
LO Lagrangian
$\mL_2^{\cO(\eta^0)}=\mL_2[\bar{u}_{L,R},\bar{h}]$.
The tree-level effective action is equal to the  action evaluated at the classical solution,
$\int {\rm d^D x} \, \,\mL^{\cO(\eta^0)}$.

\subsection{$\cO(\eta^1)$ fluctuations: EoM}

The background field configurations
correspond to the solutions of the classical equations of motion (EoM),
defined by the requirement that the linear term,
\bear
\mL_2^{\cO(\eta^1)}  &=& \Frac{v}{2} \bra \Delta \,\,\,
\left(   \nabla^\mu(\mF_C u_\mu)\, +\, 2 \mF_C J_P \right) \ket
\,\, +\,\,   v\,\epsilon\, \left( \Frac{1}{4} \mF_C' \bra u_\mu u^\mu\ket
-  \Frac{\partial^2 h}{v} -  V' - \bra J_S'\ket
\right)\, ,
\eear
vanishes for arbitrary $\vec{\eta}^T=(\Delta^a,\epsilon)$. This yields the EoM,
\bear
\nabla^\mu u_\mu &=& -2 \mF_C^{-1} J_{P}   - u_\mu \partial^\mu(\ln\mF_C)
\, ,\qquad \qquad
\Frac{\partial^2 h}{v} = \Frac{1}{4} \mF_C'   \, \bra u_\mu u^\mu \ket
   - V'
 -\,\bra  J_{S}'\ket \, .
\label{eq.EoM}
\eear
Here and  in the following, we abuse of the notation by writing the
background fields $\bar{u}_{L,R}$ and $\bar{h}$ as ${u}_{L,R}$ and ${h}$ for
conciseness.

\subsection{$\cO(\eta^2)$ fluctuations: 1-loop corrections }

The $\cO(\eta^2)$ term of the expansion of $\mL_2$ reads
\bear
\mL^{\cO(\Delta^2)} &=&
-\, \Frac{1}{4} \bra \Delta \nabla^2 \Delta\ket
\,  + \, \Frac{1}{16}\bra  [u_\mu,\Delta] \, [u^\mu, \Delta] \ket
\nn\\
&&
\hspace*{-1.7cm}+  \left[  \Frac{  \mF_C^{-\frac{1}{2}} \mK}{8}
\left(\Frac{\partial^2 h}{v}\right)
+ \Frac{\Omega}{16} \left(\Frac{\partial_\mu h}{v}\right)^2 \right]
   \bra \Delta^2 \ket
+ \Frac{1}{2\mF_C} \bra \Delta^2 J_S\ket
 ,
\nn\\
\mL^{\cO(\epsilon^2)} &= &
-\Frac{1}{2} \epsilon \left[    \partial^2
 - \Frac{1}{4} \mF_C''    \bra u_\mu u^\mu \ket
+     V''
+ \bra J_S''\ket
\right]\, \epsilon\, ,
\nn\\
\mL^{\cO(\epsilon\Delta)} &=&
  \,- \, \Frac{1}{2}   \epsilon  \mF_C'
\, \bra    u_\mu \nabla^\mu (\mF_C^{-\frac{1}{2}} \Delta) \ket
+ \mF_C^{-\frac{1}{2}} \epsilon \bra \Delta J_P'\ket
\, ,
\eear
in terms of $\mK= \mF_C^{-1/2} \mF_C'$ and $\Omega=2  \mF_C''/\mF_C - (\mF_C'/\mF_C)^2$.
Through a proper definition of the
differential operator $d_\mu \vec{\eta}=\partial_\mu \vec{\eta} + Y_\mu
\vec{\eta}$, one can rewrite $\mL_2^{\cO(\eta^2)}$ in the canonical form
\bear
\mL_{2}^{\cO(\eta^2)}   &=& - \Frac{1}{2} \vec{\eta}^T\, (d_\mu d^\mu +\Lambda) \vec{\eta}\, ,
\label{eq.Lagr-quad-fluctuation}
\eear
where $d_\mu$ and $\Lambda$ depend on ${h}$, ${u}_{L,R}$
and on the gauge boson and fermion fields (see App.~A in Ref.~\cite{Guo:2015}).
They scale according to the chiral counting as
\bear
d_\mu \,\,\, \sim\,\,\, \cO(p)\,,  \qquad\qquad \Lambda\,\,\, \sim\,\,\, \cO(p^2)\, .
\eear

The quadratic form~(\ref{eq.Lagr-quad-fluctuation}) yields a Gaussian integration over $\vec{\eta}$  in
the path-integral, which gives the one-loop contribution to the effective action,
\bear
S^{1\ell} &=& \Frac{i}{2} {\rm tr}\, \log\left(d_\mu d^\mu +\Lambda\right)\, .
\eear
where tr  stands for the full trace of the operator,
including the trace in the adjoint representation of the flavour space and
that in the coordinate space.

The computation of the full one-loop effective action is in general a difficult task.
However, it is easier
to extract its UV-divergent part.~\footnote{
It is also sometimes possible to compute the effective potential.
See for instance~\cite{Meissner:2015}.
}
For this, we first transform the trace of the log in configuration space into an integral of an exponential
by means of the Schwinger-DeWitt proper-time representation
embedded in the heat-kernel expansion~\cite{heat-kernel}:
\bear
\bra x|\log\left(d_\mu d^\mu +\Lambda\right)|x\ket &=&
\,-\, \Int_0^\infty \Frac{d\tau}{\tau}   \, \underbrace{ \bra x| e^{-\tau(d_\mu d^\mu +\Lambda)} |x\ket  }_{\equiv H(x,\tau)}
\,\,\, +\,\,\, C
\nn\\
&&\stackrel{\rm dim-reg}{=}\quad
 -\, \Frac{i}{(4\pi)^{D/2} }\, \sum_{n=0}^\infty m^{D-2n} \, \,\Gamma\left(n-\frac{D}{2}\right)\,\,  a_n(x)
\,\,\, +\,\,\, C  ,
\eear
where we obtain the expansion in term of local operators of increasing
dimension given by the
Seeley-DeWitt coefficients $a_n(x)$ and the potential UV divergences
are contained in the Gamma function
for $D\to 4$.
Only the terms of the series with $2n\leq D$ are divergent and they have their origin
on the short-distance part of the integral, this is, in its lower limit $\tau\to 0$
(the integration variable $\tau$ has dimensions of length-square in natural units).
The (infinite) constant $C$ is irrelevant here.  In the second line we have used
the Fourier decomposition of the heat-kernel in momentum space,
\bear
H(x,\tau) = \bra x| e^{-\tau (d_\mu d^\mu +\Lambda)} |x\ket
 =  \Int\Frac{{\rm d^D p}}{(2\pi)^D}
e^{-ipx} e^{-\tau (d_\mu d^\mu +\Lambda)} e^{ipx}
  =
\Frac{i e^{-\tau m^2}}{(4\pi \tau)^{D/2}}\sum_{n=0}^\infty a_n(x)\, \tau^n\, ,
\eear
where the coefficients $a_n(x)$ are extracted by expanding the interaction part of $(d_\mu d^\mu +\Lambda)$
in the exponential in powers of $\tau$
and performing the integral of each corresponding term in dimensional regularization.
In our case,
the residue of the $(D-4)^{-1}$ pole is given by the trace
Tr$\{a_2(x)\}$~\cite{Guo:2015,heat-kernel}:~\footnote{
The IR regulator $m$ of the heat-kernel integral can be made arbitrary small
and hence the term Tr$\{a_1(x)\}=\, -$Tr$\{\Lambda\}$
does not contribute to the UV divergent part;
note that the particle masses are accounted as a perturbation,
i.e., within $\Lambda(x)$.
}
\bear
S^{1\ell}    
&=& \,-\, \divergence \, \Int {\rm d^D x} \,\,\,   {\rm Tr}\{ a_2(x)\}   \, +\, {\rm finite}
\nn\\
&&=\,-\, \divergence \, \Int {\rm d^D x} \,\,\,   {\rm Tr}
\left\{ \Frac{1}{12}  [d_\mu,d_\nu] \, [d^\mu , d^\nu]
+\Frac{1}{2}\Lambda^2
\right\}  \, +\, {\rm finite}
\nn\\
&&=
\,-\, \divergence \, \Int {\rm d^D x} \,\,\, \sum_k \, \Gamma_k \, \mO_k \,
+\, {\rm finite}\, ,
\label{eq.1loop-div}
\eear
where Tr  refers to the trace over the $4\times 4$
operators that acted on
the fluctuation vector $\vec{\eta}$ in Eq.~(\ref{eq.Lagr-quad-fluctuation}).
The UV-divergence is determined by the non-derivative quadratic fluctuation $\Lambda$
and the differential operator $d_\mu$ through
$[d_\mu, d_\nu]=Y_{\mu\nu} =\partial_\mu Y_\nu -\partial_\nu Y_\mu
+[Y_\mu,Y_\nu]$,with both $\Lambda,\, Y_{\mu\nu}\,\sim\, \cO(p^2)$.
By looking at the second line of~(\ref{eq.1loop-div})
it is then clear that the UV-divergences that
appear from one-loop diagrams with $\mL_2$ vertices are $\cO(p^4)$ and
that they require counterterms of that order
to be cancelled out.
However, as some of this $p$ factors are actually constants (e.g., Higgs masses $m_h$) the structure of
the operators $\mO_k$ resembles that of other operators already present in $\mL_2$.
In general, the one-loop UV-divergences in the effective action will have a local structure and can be written
in terms of the basis of operators of chiral dimension $p^2$, $p^4$,
etc.~\cite{Longhitano:1980iz,ECLh-other,EW-chiral-counting,ECLh-Gavela},
\bear
S^{1\ell ,\, \infty} &=&  \Int{\rm d^D x}\, \bigg(
\mL_2^{1\ell ,\, \infty}  \,+\,
\mL_4^{1\ell ,\, \infty} \, +\, ... \bigg)\, .
\eear

Let us summarize the results for the NLO UV-divergences from $h$ and $\omega^a$ loops:
\begin{itemize}

\item
{\bf UV divergences with the structure of the $\mL_2$ operators in Eq.~(\ref{eq.L2}):}
\bear
\mL_2^{1\ell,\, \infty} &=&
  - \divergence
\bigg\{
\Frac{1}{8}\left[ \Frac{ \mF_C' V'}{\mF_C}(4-\mK^2)
- \mF_C\Omega V''\right] \bra u_\mu u^\mu\ket
\nn\\
&&  -\, \frac{3\mF_C' V' \Omega}{8\mF_C}
\left(\Frac{\partial_\mu h}{v}\right)^2
 +\left[\Frac{1}{2}\left( V''\right)^2
+ \Frac{3 \mK^2}{8  \mF_C    }  \left(V' \right)^2 \right]
\nn\\
&&
+ \left( V'' \bra  J_{S}''     \ket
 - \Frac{3 \mF_C' V' }{2\mF_C} \bra \Gamma_{S} \ket  \right)\bigg\}
\, ,
\label{eq.L2-div}
\eear
where
$\Gamma_{S}= \mF_C^{-1} ( J_{S} - \mF_C' J_{S}' /2)$ is an $\cO(p^2)$ tensor.
These UV divergences are cancelled out through the renormalization of
various parts
of $\mL_2$:
the couplings in the $\mF_C$ term (1st line);
the Higgs kinetic term (1st term in 2nd line),
which requires a NLO Higgs field redefinition;
the coefficients of the Higgs potential, e.g. the
Higgs mass (2nd bracket in 2nd line); the Yukawa term couplings in $\hat{Y}$ (3rd line).

\item
{\bf UV divergences with the structure of the $\mL_4$ operators:}
the $\mL_4^{1\ell,\infty}$ terms are further classified here into two types,
according to whether they include fermion
fields or not.

\begin{enumerate}

\item
{\bf Fermionic operators $\mL_4^{1\ell,\infty}|_{\rm Fer}$: }
\bear
&& \mL_4^{\rm 1\ell,\, \infty}|_{\rm Fer} =
  - \divergence
\bigg\{
\bra   \left(\Frac{\mK^2}{4}-1\right)  \Gamma_{S}
 - \Frac{\mF_C \Omega }{8}    J_{S}''\ket \, \bra u^\mu u_\mu \ket
\nn\\
&&
+ \Frac{3}{4} \Omega  \bra \Gamma_{S}\ket \, \left(\Frac{\partial_\mu
h}{v}\right)^2
+ \Frac{1}{2} \Omega  \bra \Gamma_{P} u^\mu \ket \,
\left(\Frac{\partial_\mu h}{v}\right)
\nn
\\
&&
 + \Frac{1}{2}  \bra {J_{S}''}\ket^2
+ \Frac{3}{2}   \bra \Gamma_S\ket^2
+ \frac1{\mF_C} \left( 2  \bra\Gamma_{P}^{\,2} \ket
-    \bra \Gamma_{P}\ket^2\right)
\,\bigg\} ,
\label{eq.fermion-div}
\eear
with  $\Gamma_P=  J_{P}' - \mF_C^{-1} \mF_C' J_{P} /2$ being an $\cO(p^2)$ tensor.

\item
{\bf Purely bosonic $\cO(p^4)$ divergences $\mL_4^{1\ell,\infty}|_{\rm Bos}$:}
This is actually the main result of our computation as these $\cO(p^4)$ operators of the effective action
can be only produced from the derivative interactions in $\mL_2$.
They spoil the renormalizability of the SM Lagrangian and lead to the appearance of
``true'' $\cO(p^4)$ UV-divergences, this is, operators with 4 covariant derivatives.~\footnote{
In the case that the covariant derivatives act on $h$ they just reduce to partial derivatives.
Note that the field-strength tensors can be always realized as the commutator of two covariant derivatives.
}
The outcome is summarized in Table~\ref{tab.div}.

\end{enumerate}

\end{itemize}

\begin{table*}[!t]
\begin{center}
\renewcommand{\arraystretch}{1.5}
\begin{tabular}{|c|c|c|c|}
\hline
 $c_k$  &
Operator ${\cal O}_k$ &
$\Gamma_k$ & $\Gamma_{k}[0]$  \\
\hline
\hline
  $c_1$   &  
$\Frac{1}{4}\bra {f}_+^{\mu\nu} {f}_{+\, \mu\nu}
- {f}_-^{\mu\nu} {f}_{-\, \mu\nu}\ket$
&  $\Frac{1}{24}(\mK^2-4)$
& $-\Frac{1}{6}(1-a^2)$
\\
\hline
   $(c_2 -c_3)$    &  
$\frac{i}{2}  \bra {f}_+^{\mu\nu} [u_\mu, u_\nu] \ket$
&  $\Frac{1}{24}(\mK^2-4)$
& $-\Frac{1}{6}(1-a^2)$
\\
\hline
  $c_4$   &   
$\bra u_\mu u_\nu\ket \, \bra u^\mu u^\nu\ket $
&  $\Frac{1}{96}(\mK^2-4)^2$
& $\Frac{1}{6}(1-a^2)^2$
\\
\hline
   $c_5$     &    
$ \bra u_\mu u^\mu\ket^2$
&  $\Frac{1}{192} (\mK^2-4)^2 + \Frac{1}{128} \mF_C^2 \Omega^2$
& $\Frac{1}{8}(a^2-b)^2 + \Frac{1}{12} (1-a^2)^2$
    \\
\hline
  $c_6$    &  
$\Frac{1}{v^2}(\partial_\mu h)(\partial^\mu h)\,\bra u_\nu u^\nu \ket$
&  $\Frac{1}{16}\Omega  (\mK^2-4) - \Frac{1}{96} \mF_C \Omega^2  $
& $-\Frac{1}{6}(a^2-b)(7a^2 -b-6)$
     \\
\hline
$c_7$     &   
$\Frac{1}{v^2}(\partial_\mu h)(\partial_\nu h) \,\bra u^\mu u^\nu \ket$
&  $\Frac{1}{24}\mF_C \Omega^2 $
& $\Frac{2}{3}(a^2-b)^2$
    \\
\hline
$c_8$      &  
$\Frac{1}{v^4}(\partial_\mu h)(\partial^\mu h)(\partial_\nu h)(\partial^\nu h)$
&  $\Frac{3}{32}\Omega^2$
& $\Frac{3}{2}(a^2-b)^2$
     \\
\hline
$c_9$    &  
   $\Frac{(\partial_\mu h)}{v}\,\bra f_-^{\mu\nu}u_\nu \ket$
&  $\Frac{1}{24} \mF_C' \Omega$
&  $-\Frac{1}{3}a (a^2-b)$
\\
\hline
  $c_{10}$    & 
$  \Frac{1}{2} \bra {f}_+^{\mu\nu} {f}_{+\, \mu\nu}
+ {f}_-^{\mu\nu} {f}_{-\, \mu\nu}\ket$
&  $-\Frac{1}{48}(\mK^2+4)$
& $-\Frac{1}{12}(1+a^2)$
\\
\hline
\end{tabular}
\end{center}
\caption{{\small
Purely bosonic operators needed for the renormalization of
the NLO effective Lagrangian $\mL_4$~\cite{Guo:2015}. In the last column, we provide
the first term $\Gamma_{k}[0]$ in the expansion of the $\Gamma_k$ in powers of
$(h/v)$ by using $\mF_C=1+2 a h/v + b h^2/v^2 +\cO(h^3)$.
The first five operators $\mO_{i}$
have the structure of the respective $a_i$ Longhitano operator~\cite{Longhitano:1980iz,Morales:94}
(with $i=1...5$). In addition, $c_6=\mF_{D7}$, $c_7=\mF_{D8}$  and $c_8=\mF_{D11}$ in the notation
of Ref.~\cite{EW-chiral-counting}. The last operator of the list,
$\mO_{10}=2\bra r_{\mu\nu} r^{\mu\nu}+\ell_{\mu\nu}\ell^{\mu\nu}\ket$, only
depends on the EW field strength tensors  and its coefficient is labeled as
$c_{10}=H_1$ in the notation of Ref.~\cite{chpt}.
In the notation of~\cite{Santos:2015} $c_{10}=\mF_2$, $c_2-c_3=\mF_3$ and  $c_k=\mF_k$ for $k\neq 2,3$.
}}
\label{tab.div}
\end{table*}

One can observe that the non-linearity of the $\mL_2$ Lagrangian
(where, in general, $h$ is not introduced via a complex double $\Phi$)
is the origin of these higher-dimension divergences.
For $\mF_C[h/v] = 1+  2 a h/v + b h^2/v^2 +...$, the combinations $\mK$ and $\Omega$ that rule the structure of
the divergences are given by
$\left( \mK^2 -4\right)\,=\, 4(a-1) \, +\, \cO(h/v)$,
$\Omega \,=\, 4(b-a^2)\,+\, \cO(h/v)$.
In particular in the linear limit $\mF_C=(1+h/v)^2$ and $\hat{Y}[h/v] = \, Y\, (1+h/v)$,
all the $\cO(p^4)$ divergences from $h$ and $\omega^a$ loops disappear,
\vspace*{-0.35cm}
\bear
\left(\mK^2-4\right)\,=\, \Omega \,=\, 0\, ,
\qquad\qquad
J_S''\,=\, \Gamma_S\,=\, \Gamma_P\,=\, 0\, ,
\eear
where the cancellation in the first identity relies only on the form of $\mF_C$, and
the second one --related to four-fermion operators in $\mL_4^{1\ell,\infty}$--
requires also the linear structure in $\hat{Y}$.

\section{Renormalization at NLO in the ECLh}

In order to have a finite 1-loop effective action
the divergences in Eq.~(\ref{eq.1loop-div}) are canceled by
the counterterms
\bear
\mL^{\rm ct}  
&=&  \sum_k \, c_k \, \mO_k\, ,
\eear
$
\mbox{such that }
\mL^{\rm ct}\, +\, \mL^{1\ell,\infty} \,=\, {\rm finite}\, ,
$
where the $\mO_k$ is the previous basis of EFT operators,
translating into the renormalization conditions
\bear
c_k = c_k^r + \divergence \, \Gamma_k\, .
\eear
This leads to the renormalization group equations (RGE)
for the $\cO(p^4)$ coupling constants,
\bear
\Frac{d c_{k,n}^r}{d\ln\mu } &=&
- \, \Frac{\Gamma_{k,n}}{16\pi^2}\, ,
\qquad
\mbox{with } \quad
\Gamma_k[h/v]=\sum_n \Frac{ \Gamma_{k,n}}{n!}  \left(\Frac{h}{v}\right)^n \, ,
\quad c_k[h/v]=\sum_n \Frac{ c_{k,n} }{n!} \left(\Frac{h}{v}\right)^n  \, .
\eear

Physically, this means that the NLO effective couplings
will appear in the amplitudes in combinations with logarithms of IR scales $p$.

\section{Conclusions}

Modifying the LO Lagrangian of our EFT action by allowing a non-linear structure
for the Higgs field
($\mF_C\neq (1+h/v)^2$) has important implications not only at lowest order but also at NLO.
Any misalignment between Higgs $h$ and Goldstone fields $\omega^a$ that does not allow
us to combine them into a complex doublet
$\Phi$ produces a whole new set of divergences absent in linear theory.
Nevertheless, it is possible to find combinations of couplings that are
renormalization group invariant (RGI).
Some examples derived in~\cite{Guo:2015}
are the couplings that determine $\gamma\gamma\to ZZ, \, Z\gamma,\, \gamma\gamma$,
$\gamma\gamma,\, Z\gamma\to h, \, hh,\, hhh...$
Some of the latter RGI relations were known from previous works
($\gamma\gamma\to ZZ$~\cite{1loop-AA-scat},
$\gamma\gamma,\,Z \gamma\to h$~\cite{1loop-AA-scat,Azatov:2013}).
In addition our result also reproduces the running found
in $WW,\, ZZ$ and $hh$ scattering~\cite{1loop-WW-scat-Dobado,1loop-WW-scat-Espriu}.

Our computation in~\cite{Guo:2015} did not address the following two issues:
on the one hand, the analysis of the additional contributions
to the Higgs potential and their phenomenological implications;
on the other hand, deviations from the linear-Higgs scenario leads to the appearance of UV-divergences and logs related to
four-fermion operators, e.g. $\bra J_{S,P}^2\ket$, which could be strongly constrained by flavour tests.
The study of the latter, together with the full NLO computation including gauge boson and
fermion loops, is postponed for future work.

\end{document}